\documentclass[12pt,leqno]{article}
\usepackage{amsmath,amsfonts}
\usepackage{graphicx}
\usepackage{amsthm,amssymb,bm}
\usepackage{setspace}

\begin{document}

\title{Searching for integrable Hamiltonian systems with Platonic symmetries}
\author{Giovanni Rastelli \\ \\ {\it Last affiliation}: Dipartimento di Matematica,\\ Universit\`a di Torino. \\ Torino, via Carlo Alberto 10, Italia.\\ \\ e-mail: giorast.giorast@alice.it }
\maketitle

\begin{abstract}
In this paper we try to find examples of integrable natural Hamiltonian systems on the sphere $\mathbb S^2$ with the symmetries of each Platonic polyhedra. Although some of these systems are known, their expression is extremely complicated; we try here to find the simplest possible expressions for this kind of dynamical systems. Even in the simplest cases it is not easy to prove their integrability  by direct computation of the first integrals, therefore, we make use of numerical methods to provide evidences of integrability; namely, by analyzing their Poincar\'e sections (surface sections). In this way we find three systems with platonic symmetries, one for each class of equivalent Platonic polyhedra: tetrahedral, exahedral-octahedral, dodecahedral-icosahedral, showing evidences of integrability. The proof of integrability and the construction of the first integrals are left for further works. As an outline of  the possible developments if the integrability of these systems will be proved,  we show how  to build  from them new integrable systems in dimension three and, from these, superintegrable systems in dimension four corresponding to superintegrable interactions among four points on a line, in analogy with the systems with dihedral symmetry treated in a previous article.  A common feature of these possibly integrable systems is, besides to the rich  symmetry group on the configuration manifold, the partition of the latter into dynamically separated regions showing a simple structure of the potential in their interior. This observation allows to conjecture integrability for a class of Hamiltonian systems in  the Euclidean spaces. 
\end{abstract}


\section{Integrable and superintegrable systems} It is not easy to include all known completely integrable systems within a concise definition or even a list; they appear in almost every field of mathematics and mathematical physics, from algebraic geometry  to supersymmetry theory touching classical and quantum mechanics, elliptic curves, minimal sufaces, number theory, Riemann surfaces etc \cite{FH}, \cite{Cal}, \cite {Hi}, \cite {IN} \cite{Temp}. In few words, by quoting F.Helein \cite{FH}: "...working on completely integrable systems is based on a contemplation of some very exceptional equations which hide a Platonic structure: although these equations do not look trivial a priori, we shall discover that they are elementary, once we understand how they are encoded in the language of symplectic geometry, Lie groups and algebraic geometry. It will turn out that this contemplation is fruitful and lead to many results"; the word "Platonic"  must be intended here in its purely philosophical meaning.  In the following we consider classical time-independent  completely integrable Hamiltonian systems. In this context, complete integrability coincides with Liouville integrability: a Hamiltonian system with $n$ degrees of freedom is Liouville integrable if it admits $n$ functionally independent and Poisson-commuting first integrals (i.e. constants of the motion), including the Hamiltonian function itself; the integral curves of the system are then determined by  quadratures \cite{Ar}. 
Finite dimensional Hamiltonian systems can admit up to $2n-1$ functionally independent first integrals. Liouville integrable Hamiltonian systems with extra independent constants of the motion are said to be superintegrable \cite{RW}. A well known example of completely integrable and  superintegrable system is the Kepler system in the plane, whose Hamiltonian in polar coordinates $(r,\psi)$ is $H=\frac 12(p_r^2+\frac 1{r^2}p_\psi^2)+\frac kr$ where $k$ is a real constant, its functionally independent first integrals are the Hamiltonian, the angular momentum  and one of the components of the Laplace vector (the other depends functionally on these three) \cite{LaLi}. The integral curves of (time independent) Hamiltonian systems belong to  the phase-space (cotangent bundle, in the language of symplectic geometry), whose dimension is twice the degrees of freedom of the Hamiltonian; the existence of $2n-1$ first integrals implies that each integral curve is geometrically determined by the one-dimensional intersection of all the $2n-1$ dimensional hypersurfaces corresponding to the constants of the motions determined by the initial conditions, then,  without need of integrating the Hamilton differential equations, even if it can be practically impossible its explicitation in that way due to the mathematical complexity of the functions involved. If the independent constants of the motion are $2n-2$, then each integral curve will stay in their  two-dimensional intersection and so on. The $n$ Poisson-commuting independent first integrals $K_i$ of a Liouville integrable Hamiltonian, under the assumption of completeness and  excluding eventual critical points of the Hamiltonian fields, generate the foliation of the phase space into the n-dimensional submanifolds determined by the intersection of the manifolds $K_i=const.$, which are diffeomorphic to $\mathbb R^k\times \mathbb T^{n-k}$, for $0\leq k \leq n$, where $\mathbb T^{n-k}$ is the $n-k$ dimensional torus (known as Liouville torus). If the Hamiltonian has two degrees of freedom, one  first integral other than H assures the Liouville integrability of H and if the submanifolds of the foliation are compact, then they are diffeomorphic to  tori \cite{Ar}. 

\section{Poincar\'e sections} 
\begin{figure}
\begin{minipage}[b]{ 440pt} 
\begin{center}
\includegraphics*[width=   220pt, height=   200pt]{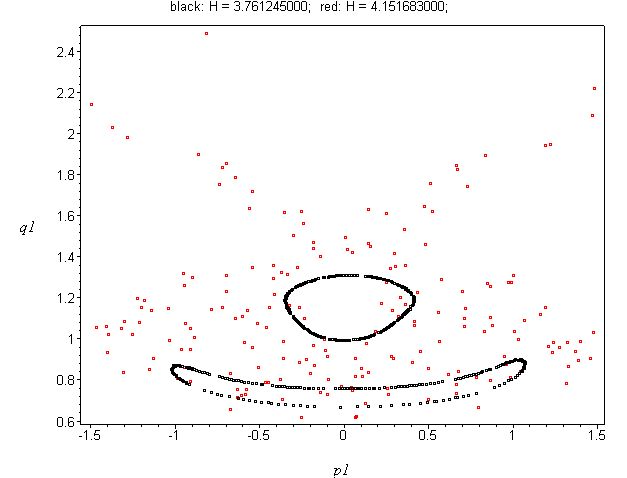}
\end{center}
\caption{{\it A cross section showing intersections of two distinct trajectories for $V_{TO}$. Black squares correspond to an orbit lying on some 2-dimensional submanifold of the phase space while the magenta ones correspond to an orbit not lying on a surface. Therefore, the system do not show  behaviour of complete integrability.}
\label{FN}}
\end{minipage}
\end{figure}

For two-dimensional completely integrable Hamiltonian systems the phase space  is foliated into two-dimensional manifolds and each integral curve of the system belongs to one and one only of the leaves of the foliation. In this case, the intersections  of any integral curve with a given plane of the phase space ("Poincar\'e sections" or "phase sections") consist  of points arranged into curves, determined by the intersection of the two-dimensional Liouville torus with the section plane. Naturally, the Liouville tori are only diffeomorphic to the standard torus, then they appear twisted and folded into the phase space so that their intersections with a section plane can produce several disconnetted closed curves.  This behaviour is a characteristic feature of two-dimensional integrable systems. Poincar\'e sections of non-integrable systems, instead, can show both  points arranged in curves or points shattered into the section plane, depending on their different initial data (Figure \ref{FN}), \cite{LL}. If one want to show the integrability of a system by means of Poincar\'e sections only, it would be necessary to produce sections for all possible sets of initial conditions all showing points forming curves; however, this is clarly impossible in practice. The only proof of the existence of first integrals, and complete integrability, is the knowledge of their mathematical expression itself and the validity of Poisson equations. Such a knowledge can be extremely hard to obtain for Hamiltonian with not simple potentials. An experimental approach is however possible: numerical integration of the integral curves of the system can be performed for generic initial data and Poincar\'e sections can be obtained. Several computer algebra systems can do the task and we used the "poincare" procedure implemented in Maple 9.5. By analyzing the Poincar\'e sections so obtained we can have an experimental evidence of integrability, in the case all sections we obtain for generic initial data show the features expected in case of integrable systems, and therefore focalize the efforts in search of a rigorous proof. We produce below some of these experimental evidences for three  potentials with the symmetries of the Platonic polyhedra.

\section{The Calogero system and dihedral symmetry} An integrable system of great importance in mathematical physics is the Calogero-Sutherland-Moser system \cite{Cal}, which is studied both in its classical and quantum form. This system, in its simplest form, describes the reciprocal interactions of three points on a line, denoted by their positions $x^i$ with respect to some origin,  with potential 
\begin{equation}
V=\frac 1{(x^1-x^2)^2}+ \frac 1{(x^2-x^3)^2}+\frac 1{(x^3-x^1)^2}.\label{Ca1}
\end{equation}
It is possible to write the Calogero system as a one-point system in $\mathbb R^3$ with cylindrical coordinates $(r,\psi,z)$ (see Section 8 and \cite{CDR}) which, because of the conservation of the momentum, reduces to a two-dimensional system in the plane  $(r,\psi)$ with potential 
\begin{equation}
V=\frac 1{r^2\sin^23\psi}.\label{Ca2}
\end{equation}
It is now evident the dihedral symmetry of the system: the plane is divided into six identical sectors where the dynamics is the same, and the symmetries of the potential are those of the hexagon (the discrete rotational symmetries of the  regular polygons and their reflectional symmetries are called dihedral symmetries). Particles moving on the plane under the potential $V$ are trapped into each sector by the infinite value of the potential attained on the boundaries of the sector. In \cite{CDR} is conjectured the super-integrability of systems in the plane  with potential 
$$
V=\frac 1{r^2 \sin^2 k \psi}
$$ 
with $k$ integer and, for odd k, the expression of the corresponding constant of the motion is given (the proof will be published soon), moreover, these systems are shown to correspond to three-body interactions among three points on a line in the same way of the Calogero system. For these systems the dihedral symmetry corresponds to the symmetry of the regular polygon with $2k$ sides. Other studies about systems with dihedral symmetry recently appeared  are \cite{Win}, \cite{Quesne}, \cite{KM3}. The hexagonal symmetry of the potential (\ref{Ca2}) is directly related to a cubic in the momenta first integral of the system  and the same holds for more  general potentials \cite{CDR}. It must be remarked that the hexagonal symmetry is completely hidden if the potential is represented as (\ref{Ca1}). The main idea of the present investigation is (try) to show that, in analogy with the Calogero system, finite symmetries of the potential, maybe together with some other ingredient, lead to  first integrals.

\section{Systems with Platonic symmetries} 
It is natural  try to generalize the previous results to the three dimensional space, then, to search for  integrable  Hamiltonian systems with potentials of the form 
\begin{equation}
V=\frac 1{ \rho^2f(\theta, \psi)}\label{V}
\end{equation}
in spherical coordinates $(\rho,\theta,\psi)$ where $f$ is a function on the sphere $\mathbb S^2$ , with polyhedral symmetry. Integrable systems of this kind will be superintegrable if embedded in $\mathbb R^4$ (with 5 independent constants of the motion) and equivalent to superintegrable interactions among four points on a line, as we show in Section 9. The resulting natural Hamiltonian system in the three-dimensional Euclidean space is sometimes referred to as "conformal mechanics"  and its construction  is generalizable to the building of  integrable and superintegrable systems from spheres $\mathbb S^{n-1}$ to $n$-dimensional Euclidean spaces  \cite{CDR}, \cite{CDR1}, \cite{HKLN} (where connections are made with Calogero system, Higgs oscillators and supersymmetric mechanics). 

We limit our investigation to potentials with the same symmetries of the five Platonic polyhedra as the simplest three-dimensional generalization of the dihedral symmetry manifested by the Calogero and the other systems seen above. There is no need to remark the relevance  of Platonic polyhedra in philosophy, arts and science, more can be found for example in \cite{Crom}. 
The Platonic polyhedra are those polyhedra whose faces are all made by the same regular polygons and whose vertices belong to a sphere. Their symmetries are the rotations and reflections leaving fixed the center of the polyhedron and making the faces to correspond, leaving in this way unaltered the appearence of the solid: the polyhedral symmetries are those rotations and reflections which leave invariant the polyhedron.  Since Euclid's times it is known that there are only five possible Platonic polyhedra: tetrahedron, exahedron (or cube), octahedron, dodecahedron and icosahedron. Actually, esahedron-octahedron and dodecahedron-icosahedron form so called dual (or reciprocal) couples (the centers of the faces of one correspond to the vertices of the other) and each couple share the same symmetries while the tetrahedron is self-dual. The same symmetries  are shared by the Archimedean polyhedra obtained from the Platonic ones, and by their duals the Catalan solids \cite{Crom}, \cite {Co}, \cite{IN}. Namely, the symmetries we are considering are the polyhedral groups denoted by $T_{12}$, $O_{24}$ and $I_{60}$ (isomorphic to  $A(4)$, $S(4)$, $A(5)$ respectively,  $A(n)$ and $S(n)$ denoting the alternate and symmetric grups of degree $n$ respectively), where the lower number is the order of the group.

An example of superintegrable system with platonic symmetry whose first integrals can be explicited is given in \cite{Cuboct}. It is obtained  from the three particle  Calogero system $D_3$ \cite{OlPer} and interpreted as potential generated by  six centers of force on the sphere, each one on the vertices of a cuboctahedron (one of the Archimedean polyhedra). It is  a superintegrable system with the symmetries of the exahedron-octahedron, and its potential on the sphere is 
\begin{eqnarray*}
V_{CO}=&\dfrac{9(8-\tan^2\theta)^2}{2(3\tan^2\theta-8+\tan^3\theta\cos 3\psi)^2}+\dfrac {12}{3\tan^2\theta-8+\tan ^3\theta \cos 3\psi}+\cr
&+\dfrac 9{4\sin^2\theta (1+\cos 6\psi)},
\end{eqnarray*}
its first integrals can be obtained by following \cite{Cuboct}. Here, the integrability of the system is derived from that of the original Calogero system $D_3$, the complexity of the potential make almost impossible any direct inquiry in that direction. We try here to build simpler integrable  systems with platonic symmetries in order to provide more suitable matter for further analysis.

Symmetry groups for Platonic polyhedra and their polynomial invariants, i.e. polynomials in CArtesian coordinates  $(x,y,z)$ left invariant by the symmetry groups of the corresponding polyhedra, are well known. Examples of applications of platonic symmetries in physics and mathematics are \cite{Be} or \cite{IN}, generalizations are currently object of investigation relatively to more general symmetry groups \cite{Iwa}. 

It can be shown that all invariant  polynomials over the reals $\mathbb R$ (or the complex $\mathbb C$) for each one of the symmetry group of the Platonic polyhedra can be written as polynomials, over the field $\mathbb R$ (or $\mathbb C$), in the variables $[U_1,U_2,U_3]$, where the $U_i$ are suitable homogeneous poynomials in Cartesian coordinates $(x,y,z)$ (an instance of the celebrated Hilbert finite base theorem) \cite{PS}, \cite{BS}. Because of their homogeneity, these polynomials,  written in spherical coordinates, can always be factorized into the form $U_i= \rho^kf_i(\theta, \psi)$ for some positive integer $k$, this correspond to some "conformal invariance" of the system \cite{HKLN}.
Therefore, the functions $f_i$ on $\mathbb S^2$ so determined carry all the polyhedral symmetries of the original polynomial, the same do $f^{-1}_i$ and any function of the form $g(\rho)f^{-1}_i$ for arbitrary functions $g(\rho)$. In analogy with the two-dimensional dihedral case, we will consider potential functions on $\mathbb S^2$ of the form $V=f^{-1}$ and in $\mathbb R^3$ of the form $W= \rho^{-2}f^{-1}$. To produce our  examples we consider some  of the  polynomials of the bases $(U_i)$  as given in \cite{PS}, due to the fact that they have possibly the simplest algebraic expression for functions with the desired symmetries.

In these bases, the lowest-order polynomial is always $x^2+y^2+z^2$ which obviously  is invariant under all rotations and reflections leaving fixed the origin,  while, 
$$
T=xyz
$$ 
for the tetrahedron, 
$$
O=x^2y^2z^2
$$ 
for the exahedron-octahedron and 
$$
I=-z(2x+z)(x^4-x^2z^2+z^4+2(x^3z-xz^3)+5(y^4-y^2z^2)+10(xy^2z-x^2y^2))
$$
for the dodecahedron-icosahedron, are characteristic of each of them. The polynomial $TO=x^2y^2+x^2z^2+y^2z^2$ is common to the bases of tetrahedral and octahedral invariant polynomials. By writing these polynomials in spherical coordinates and by factorizing out the radial terms, we obtain, as described above,  the functions $f_i$ for the polynomials $T$, $O$ and $I$ and from $f_i^{-1}$ the following "platonic"  potentials  on the sphere:
$$
V_T=(\sin^2\theta\cos\theta\cos\psi\sin\psi)^{-1}
$$ 
with tetrahedral symmetry, 

$$
V_O=V_T^2
$$ 
with cubic-octahedral symmetry, and
\begin{eqnarray*}
V_I=&-\cos^{-1}\theta [ \cos^5\theta-5\sin^2\theta\cos^3\theta+5\sin^4\theta\cos\theta+\cr
&\sin^5\theta( 32\cos\psi\sin^4\psi-24\cos\psi\sin^2\psi+2\cos\psi)]^{-1},
\end{eqnarray*}
with dodecahedral-icosahedral symmetry.

Still, it is not easy to find directly the expressions of first integrals of these systems, even by using computer algebra methods as we did in the lower-dimensional case \cite{CDR}. Therefore, we integrate numerically the natural Hamiltonian systems with the three potentials of above on the sphere and analyze their Poincar\'e sections.

Poincar\'e sections are meaningful only if orbits belong to  a compact submanifold of the phase space, in this case each orbit, numerically computed, winds itself in general several times around the Liouville torus and can produce several points on a suitably chosen section plane. This happens in our case  in the regions of $\mathbb S^2$ where the potential is positive. In the remaining regions the potential generates a force pulling the particle towards the borders of the region and the orbit crosses the section planes only few times. The structure of the Hamiltonian in the negative-potential regions could be studied by considering $-V$ instead of $V$, which of course are different systems. 

\begin{figure}
\begin{minipage}[b]{ 440pt} 
\begin{center}
\includegraphics*[width=   420pt, height=   200pt]{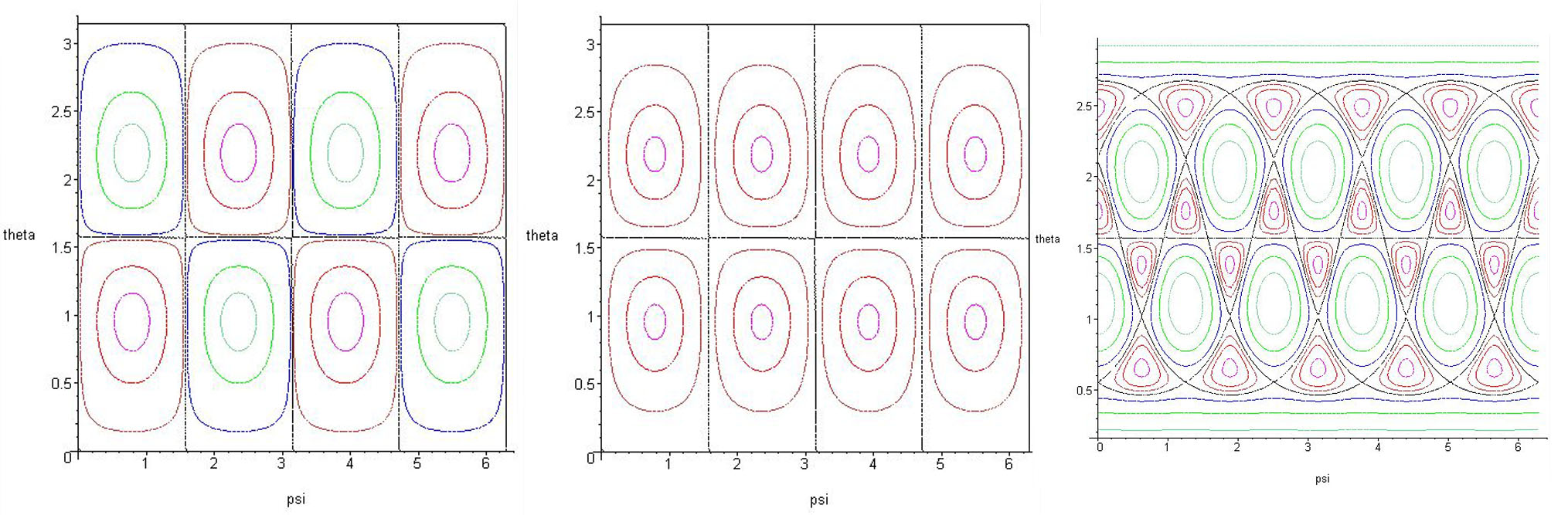}
\end{center}
\caption{{\it  For the potentials $V_T$, $V_O$ and $V_I$ with tetrahedral, cubic-octahedral and dodecahedral-icosahedral symmetry respectively, some isopotential lines are drawn on the sphere $\mathbb S^2$. Aquamarine-green-blue denote decreasing negative and magenta-orange-red increasing positive values of the potential. Black lines denote infinite values of the potential, therefore, they determine regions of the sphere where the motion is confined. Regions with identical behaviour of the potential correspond under the symmetry group of the associated polyhedron. The isopotential lines show the simple structure of the potential in each region of the sphere bounded by the black lines.}
\label{Ft0}}
\end{minipage}
\end{figure}

It is remarkable that in all these examples the sphere is partitioned by the lines of the zeroes of the $f_i$ into separated regions. Because on these lines the value of $V$ goes to infinity, particles moving under the potential $V$ cannot cross the borders of these regions. In these regions, the potential admits just one critical  point which is either a maximum or a minimum (Figure \ref{Ft0}), therefore, the structure of the dynamics is essentially simple: there are on the sphere dynamically separated regions each one with some simple dynamical structure.    The same happens for the Calogero system considered above: the sphere $\mathbb S^1$ is partitioned into six regions by the infinities of the potential, regions equivalent under the symmetries of the hexagon. 

\begin{figure}
\begin{minipage}[b]{ 440pt} 
\begin{center}
\includegraphics*[width=   420pt, height=   200pt]{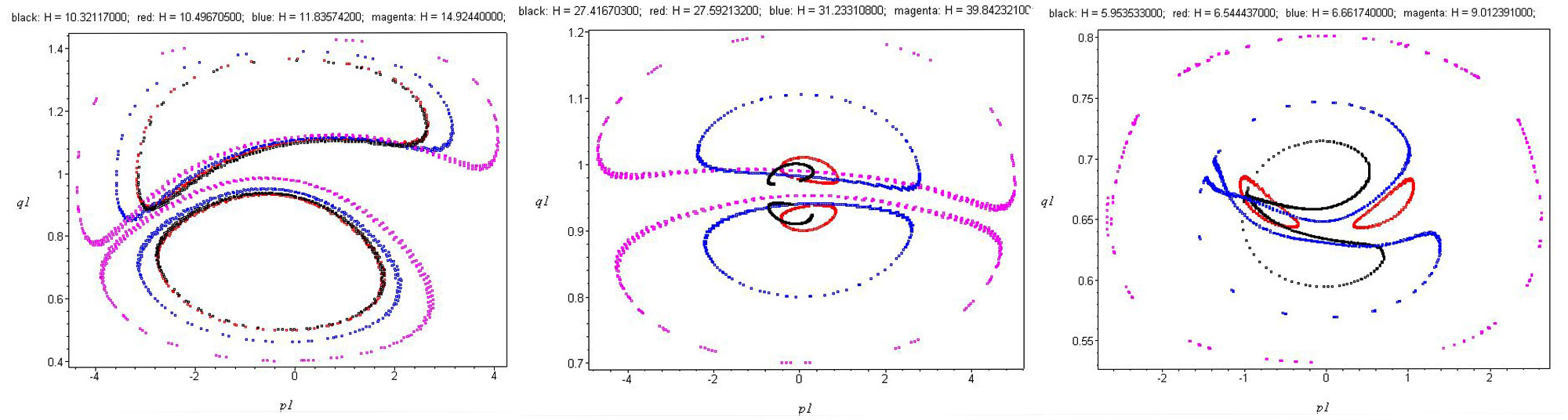}
\end{center}
\caption{{\it Are shown here examples of Poincar\'e sections, with $q_1=\theta$, $p_1=p_\theta$ the momentum conjugate to $\theta$, of integral curves of  natural systems on the sphere with potential, from left to right, $V_T$, $V_O$ and $V_I$ with four distinct initial condition sets, distinct by their colors. After some minimal level of accuracy in the numerical integration of the integral curves, their intersection points with a plane $(p_1,q^1$) in the phase space describe closed curves, a behaviour compatible with complete integrability.}\label{Ft1}}
\end{minipage}
\end{figure}

Possibly as a consequence of that additional structure, the corresponding Hamiltonian systems appear to be Liouville integrable: the computation of Poincar\'e sections for several randomly chosen initial conditions in the regions where the potential is positive shows in fact always the  curvilinear features of the intersections of the integral curves with the section planes, giving evidence of an independent first integral $K(p_\theta,p_\psi,\theta,\psi)$ at least for each system (Figure \ref{Ft1}).  
This happens also for the systems with potential $-V$ in the regions where $V$ is negative. The simple potential in each region can allow the existence of a local first integral which is extended on the other regions corresponding under the platonic symmetry group. 
The same approach with potentials on the sphere obtained from the $f_i$ instead of the $f_i^{-1}$ of above do not show yet signs of integrability. 
Some computations made with the potential obtained in a similar way from the third polynomial of the icosahedral base $(U_i)$ given in \cite{PS} indicate its integrability, however, much less sections have been considered in this case.

Not all the potentials  with platonic symmetry obtained from the polynomials $U_i$ seem to be integrable; as an example consider 
$$
V_{TO}=\sin^{-2}\theta[\cos^2\psi-\cos^4\psi-\cos^2\psi\cos^2\theta+\cos^4\psi\cos^2\theta+\cos^2\theta)]^{-1},
$$
obtained from the polynomial $TO$ of above, which shows six  isolated points of maximum with infinite value there.  In this case the sphere is not partitioned into regions with a simple behaviourhed potential. Neither all integrable potentials with platonic symmetries partition the sphere into dynamically separated regions with a simple potential within,  for example, $V_{CO}$. 

The integrable dynamics just analyzed on the sphere can be extended to the three-dimensional space with the potentials $W=\rho^{-2}V$. The new natural Hamiltonian systems are integrable in $\mathbb R^3$ if the original ones on the sphere are, and  their embedding in dimension 4 is superintegrable as we prove in Sections 7 and 8. An harmonic term proportional to $ \rho^2$ can be added to each one of the previous potentials in $\mathbb R^3$ keeping the integrability and allowing for finite trajectories of the systems. Then, integrability can be studied  by analyzing the orbit structure of the system by following the approach developed in dimension 2 by \cite{Win}. 
Even if we do not know yet the expression of the possible first integrals, likely they are polynomial in the momenta.

In Sections 8 and 9  we see as Integrable systems on $\mathbb S^2$ lead to superintegrable 4-points systems on a line, in the same way integrable systems on $\mathbb S^{n-2}$ can be interpreted as superintegrable $n$-body systems on a line \cite{CDR}. In this perspective could be interesting to analyze higher dimensional  polytopes (by using a Coxeter's expression, a polytope is the general term of the sequence: point, segment, polygon, polyhedron,...),  in Euclidean or non Euclidean spaces \cite{Co}.

\section{Remarks} The partition of the configuration manifold into "simple" dynamically separated regions united to some  suitably rich symmetry group on the same manifold seems  a good indicator of integrable potentials. For example,  on the Euclidean plane potentials of the form 
$$
V_1=(\sin^a hx \cos^b ky)^{-1},
$$
with $a,b,h,k$ positive integers and $(x,y)$ cartesian coordinates, partition by their lines of infinities the plane into rectangular regions where the potential has only one point of maximum or minimum and a simple behaviour. In each region of the partition the Poincar\'e sections show a dynamics compatible with integrability. On the plane, these systems admit translational symmetries along axes $x$ and $y$ of magnitude $\frac {s_x}h\pi$ (resp $\frac {s_y}k\pi$), where $s_x$ ($s_y$) are  even integers for $a$ (resp. $b$) odd and any integer otherwise, all generated by a $\frac \pi h$ ($\frac {2\pi}h$ for odd $a$) translation in the $x$ direction and by a $\frac \pi k$ ($\frac {2\pi}k$ for odd $b$) translation in the $y$ direction. The lines of reflectional symmetry have equations $x=\frac{s_x}{2h}\pi$ and $y=\frac{s_y+1}{2k}\pi$.
It seems likely that integrability holds also for the three-dimensional natural Hamiltonian systems in the Euclidean space with potential of type 
$$
W_1=(\sin^a hx \sin^b ky \sin^c lz)^{-1}
$$
with $c$ and $l$ positive integers,  as well as for the generalization of these systems to higher dimensions, where again the space is honeycombed into dynamically separated cells with simple potentials in each of them and the symmetry group is generated by the elementary translations of magnitude $\frac \pi h$, $\frac \pi k$, $\frac \pi l$ (twice in the cases of odd $a$, $b$, $c$ respectively) in the three coordinate directions, plus the obvious reflectional symmetries. In all these cases, assuming they are integrable, the integrability in each cell, which is due to some local first integral, is extended to the whole configuration manifold by the symmetry group on it and the local first integrals become a global one. 
 
Potentials of  the form  
$$
V_2=(\sin^a h\theta \sin^b k\psi)^{-1}
$$
instead,  have polyhedral but not platonic symmetry and   partition the sphere into a meridian-parallel web of "simple" dynamically separated regions by their lines of infinite value. 
In this case, the  rotations of the dihedral symmetry group on the configuration manifold are generated by the  rotation around the poles of amplitude $\frac \pi k$ for $b$ even and $\frac \pi k$ for odd $b$, a dihedral group.  The planes of reflectional  symmetry are defined by the integer multiples of $\psi=\frac \pi k$ for $b$ even and $\psi=2\frac \pi k$ for odd $b$. There is the additional reflectional symmetry with respect to the equatorial plane when $a$ is even or $h$ is odd. 
While appearing in general integrable, in some cases their  Poincar\'e sections  seem to show integrable behaviour only in  the  regions  which are not neighbouring to the poles, for example for $a=4$, $h=2$, $b=1$, $k=3$.

A behaviour analogous to the systems with potentials of type $V_2$ is shown  in the Euclidean plane by the potentials of type 
$$
V_3=(\sin^a hr \cos^b k\psi)^{-1}
$$
with, mutatis mutandis, all possible permutations of $\sin$ and $\cos$ functions, where $(r,\theta)$ are polar coordinates. In the regions neighbouring  to the origin they seem do not  to be always integrable, differently from the others regions, for example when $a=b=2$, $h=k=1$. Here, an evident dihedral symmetry group exists. In all these cases, some kind of local first integral could exist on the quadrangular regions only, but not on the triangular ones neighbouring to the origin. A local integral is extended by the symmetries of the configuration manifold to a semiglobal first integral on the whole manifold minus the regions surrounding the poles of the coordinates: the whole plane minus a disc centered on the origin of the coordinates. Due to the dynamical separation of each region, systems of this kind could be considered completely integrable as far as initial conditions are not chosen to be into the "bad" regions around the poles. 

Differently from these last examples, the  Calogero system considered above, the systems in \cite{CDR} with dihedral symmetry and the particular case of $V_2$ with $a=2$, $h=1$ on the sphere, admit the additional symmetry of separability. Indeed, in all these cases the differential equations of the dynamics are separated for the two coordinates thanks to a quadratic first integral (see \cite{CDR} for the first two cases).  On the sphere, the natural Hamiltonian with potential 
$$
V_4=\frac{F(\psi)}{\sin ^2\theta},
$$
for any function $F(\psi)$, admits the quadratic first integral $p_\psi^2+2F(\psi)$ and is therefore completely integrable and separable as in the previous cases. Symmetries like the latter are sometimes called "hidden", because not immediately recognizable from the expression of the potential, as instead is for example the central symmetry for the Kepler system which is associated with the conservation of the angular momentum. In some cases, hidden symmetries can be unveiled by much more evident ones. It is the case of the three body Calogero system shown above where the hidden symmetry is a third-order polynomial in the momenta, unveiled by the hexagonal symmetry of the potential written in the center of mass frame. 
Remarkably, in \cite{KM2} the three-dimensional system with Hamiltonian
$$
H=p_\rho^2+\frac {p_\theta^2}{\rho^2}+\frac{p_\psi^2}{\rho^2\sin^2 h\theta}+\frac \alpha \rho+\frac 1{\rho^2}\left(\frac {\beta_1}{\cos^2 h\theta}+\frac{\beta_2}{\sin^2 h\theta \cos^2 k\psi}+\frac{\beta_3}{\sin^2 h\theta \sin^2 k\psi} \right)
$$
where $\alpha$, $\beta_i$ are real parameters, is shown to be maximally superintegrable for all  $h,k$  rationals. If reduced to the submanifolds $\rho=const.$, for  $\alpha=\beta_1=0$ and $h=1$  the system becomes an  instance of the natural Hamiltonian on the sphere with potential $V_4$ and dihedral symmetry.

We think to have provided here several evidences for the presence of hidden symmetries connected both with platonic symmetries of the potential and with the partition of the configuration manifold into simple dynamically separated regions. Anyway, all the examples of above are just hints to further inquiries and we do not pretend to give here a detailed description of their behaviour.

\section{Quantization} The classical systems here considered can be transformed into quantum mechanical ones by standard quantization techniques. Indeed, the Calogero system is in origin quantistic. Instead of Hamiltonians we have in this case Schr\"odinger operators on the sphere or on the Euclidean three-dimensional space, instead of Poisson commuting quadratic in the momenta first integrals, commuting second order differential operators. More subtle is the quantization of higher-order first integrals \cite{DV} or the definition of quantum superintegrability \cite{WSi}. We do not consider further here the quantum version of our systems.

\section{First integrals in $\mathbb R^3$} The procedure described below applies to every Hamiltonian integrable system on the sphere with potential of the form (\ref{V}), in particular to the systems described in Section 4, provided they admit a first integral and are, consequently, completely integrable. The natural Hamiltonian on $\mathbb S^2$ is
$$
H_1=\frac 12(p_\theta^2+\frac 1{\sin^2 \theta}p_\psi^2)+V(\theta,\psi)
$$
and our platonic systems correspond to the case when $V$ is either $V_T$, $V_O$ or $V_I$. Let us call $H_2(\theta,\psi,p_\theta,p_\psi)$ an independent  first integral of $H_1$, (the unknown first integral inferred from the structure of the Poincar\'e sections in our case). The system is then Liouville integrable.
Let, in spherical coordinates $(\rho,\theta,\psi)$ of $\mathbb R^3$,
$$
H_3=\frac 12(p_ \rho^2+\frac 2{ \rho^2}H_1)+\frac k2 \rho^2,
$$
the natural Hamiltonian with potential $W=\frac 1{ \rho^2}V+\frac k2  \rho^2$ where the original potentials on the sphere are modified by a harmonic term with parameter $k\in {\mathbb R}^+$. The system determined by $H_3$ is completely integrable if $H_1$ is. Indeed, the three functions $H_3$, $H_1$ and $H_2$ are functionally independent and all in involution if $H_1$ and $H_2$ are. The equations of the motion are partially separated into equations in $\theta,\psi$, for the dynamics projected onto the sphere of fixed radius $\rho_0$ (the dynamics of $H_1$), and
$$
p_\rho^2=2h_3-k \rho^2-\frac 2{ \rho^2}h_2
$$
where $h_i$ are the values taken by $H_i$ in the given initial conditions.

\section{First integrals in $\mathbb R^4$.} The Hamiltonian $H_3$ of above can be extended to ${\mathbb R}_4$ \cite{CDR}, \cite{HKLN} by introducing cylindrical coordinates $(u,\rho,\theta,\psi)$ and obtaining the natural Hamiltonian
$$
H_4=\frac 12(p_u^2+2H_3)
$$
which admits the trivial first integral 
$$
H_5=p_u^2
$$
and, if $k=0$, the less trivial
$$
H_6=\frac 12(up_\rho-\rho p_u)^2+\frac {u^2}{ \rho^2}H_1,
$$
in these coordinates, $\rho$ is the distance from the axis of the three-dimensional cylindrical hypersurfaces. The functions $H_1$, $H_2$, $H_4$ and $H_5$ are still independent and in involution with each-other, while $H_6$, if $k=0$, is in involution with  $H_1$, $H_2$, $H_4$ and independent from  $H_1$, $H_2$, $H_4$ and $H_5$.
Therefore, the system of Hamiltonian $H_4$ is (minimally) superintegrable in $\mathbb R^4$.  Other examples of this kind of extension, applied to superintegrable Evans systems, are given in \cite{CDR1}.

\section{Four points on a line}  The system of above can be interpreted as a natural Hamiltonian system describing reciprocal four-body interactions among four points on a line by the change of variables
\begin{eqnarray*}
u^j&=&\frac 1{\sqrt{j(j+1)}}(x^1+\ldots +x^j-jx^{j+1}) \quad j=1\ldots 3\\
u^4&=&\frac 1{\sqrt{4}}(x^1+\ldots+x^4),
\end{eqnarray*}
where $u^1=x$, $u^2=y$, $u^3=z$, $u^4=u$ and $x^i$ denote the positions of four points on a line whose dynamics is still described by the Hamiltonian $H_4$ and keep the same integrals of the motion. The integral $H_5$ correspond to the conservation of the momentum of the system of four bodies. Evidently, the previous identification between one-point systems in $\mathbb R^4$ and four body systems on $\mathbb R$ is essentially unaffected by phase shifts in $\theta$ and $\psi$, namely transformations $\theta\rightarrow \theta+\theta_0$ and $\psi\rightarrow \psi+\psi_0$, because such phase shifts do not modify the dynamics of the four points but only their reciprocal position on the line. An example of the equivalence just remarked is given in \cite{CDR} between the Calogero and Wolfes potentials. The procedure can be extended to $n$ points on a line \cite{CDR1}.



\section{Numerical procedures} The Poincar\'e sections of this paper have been obtained by using the "poincare" procedure implemented in Maple 9.5, which is a fourth-order Runge-Kutta algorithm. For each section the integral curves have been numerically integrated for a Hamiltonian parameter $t$ ranging typically between $\pm 100$ or $\pm 50$ and  several hundreds of crossing points are obtained. The section planes parallel to $(q^1,p_1)$ or $(p_1,p_2)$ have been selected in order to show intersections for all the four sets of initial conditions and the choice is not particularly relevant regarding the shape of the sections. The procedure "poincare" automatically check the maximal percentage deviation from the initial value of the Hamiltonian along the computed integral curve, giving in this way some measure of the confidence of the integration. If the percentage deviation exceed some given maximum, the accuracy of the integration can be increased, for example, by refining the discretization of the $t$ interval. In our computations we allowed deviations up to $1\times10^{-3}\%$ of $H$, even if in most of the graphics the deviation is typically $1000$ times smaller. Point crossing the section plane are assumed do not describe curves if, after further refinements of the $t$ discretization, they keep that behaviour. Obviously, the  curves in the section planes are obtained after some minimal discretization and are stable for finer discretizations. The practical minimum of the discretization step on our computer for $t$ in the intervals of above  is around $0.002$.  

\section{Conclusions.}  In this paper we produce evidences of complete integrability for two dimensional Hamiltonian systems on the sphere with the symmetries of the platonic polyhedra. By assuming the effective existence of a first integral, we show how to extend the systems from the sphere to an integrable system in the three-dimensional space and how to build superintegrable systems in dimension four corresponding to four-body interactions on a line. Even if no explicit first integral of the platonic systems on the sphere is obtained, its existence seems more than probable. The  integrability of these systems seems ascribable to  the presence of a partition of the configuration manifold into dynamically separated regions, each one with a simple structure of the potential allowing a local first integral, and of a symmetry group allowing the extension of the local first integrals  into a global one on the whole  configuration manifold. The conjecture is extended to a class of similar natural Hamiltonians in Euclidean spaces. If the extension fails in some of the regions, the system can be considered integrable in a semiglobal sense. The approach seems  fertile for further inquiries: first, towards a determination of the first intergrals, second, towards the extension of the approach to higher dimensional polytopes and to non-euclidean spaces.

\section{Acknowledgements} I am grateful to Claudia Chanu and Luca Degiovanni for several fruitful discussions about the subject of this article.

\newpage

\begin{figure}
\begin{minipage}[b]{ 440pt} 
\begin{center}
\includegraphics*[width=   440pt, height=   200pt]{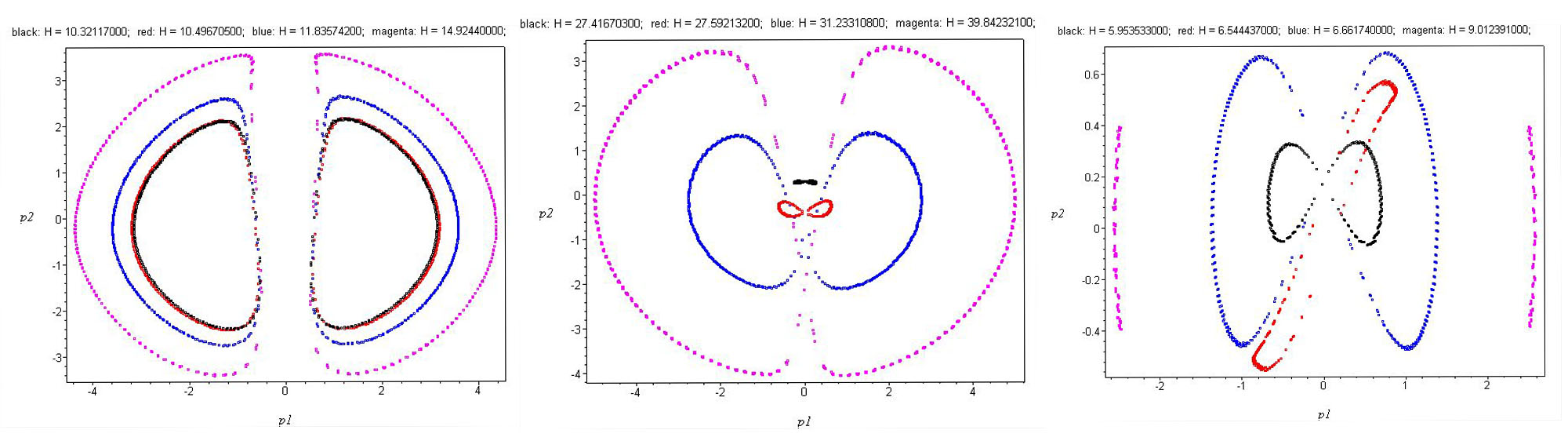}
\end{center}
\caption{{\it Poincar\'e sections in planes parallel to $(p_1=p_\theta,p_2=p_\psi)$ for $V_T$, $V_O$, $V_I$ respectively, with the same initial conditions of Figure \ref{Ft1}. }
\label{Fp1p2}}
\end{minipage}
\end{figure}

\end{document}